\RequirePackage[2020-02-02]{latexrelease}

\documentclass[aps,prb,twocolumn,superscriptaddress,longbibliography]{revtex4-1}

\usepackage[colorlinks=true,citecolor=blue]{hyperref} 
\usepackage{graphicx}
\usepackage{bm}
\usepackage{amsmath}
\usepackage{amssymb}

\DeclareGraphicsExtensions{.png,.jpg,.eps}
\usepackage{xcolor}

\newcommand{\beq}{\begin{eqnarray}}
\newcommand{\eeq}{\end{eqnarray}}
\begin{document}

\author{Guangze Chen}
\affiliation{Department of Applied Physics, Aalto University, 02150 Espoo, Finland}

\author{Malte R\"{o}sner}
\affiliation{
Institute for Molecules and Materials, Radboud University, NL-6525 AJ Nijmegen, The Netherlands}

\author{Jose L. Lado}
\affiliation{Department of Applied Physics, Aalto University, 02150 Espoo, Finland}


\title{Controlling magnetic frustration in 1T-TaS$_2$ via Coulomb engineered long-range interactions}

\begin{abstract}
Magnetic frustrations in two-dimensional materials provide a rich playground to engineer unconventional phenomena.  However, despite intense efforts, a realization of tunable frustrated magnetic order in two-dimensional materials remains an open challenge. Here we propose Coulomb engineering as a versatile strategy to tailor magnetic ground states in layered materials. Using the frustrated van der Waals monolayer 1T-TaS$_2$ as an example, we show how long-range Coulomb interactions renormalize the low energy nearly flat band structure, leading to a Heisenberg model which depends on the Coulomb interactions. Based on this, we show that superexchange couplings in the material can be precisely tailored by means of environmental dielectric screening, ultimately allowing to externally drive the material towards a tunable frustrated regime. Our results put forward Coulomb engineering as a powerful tool to manipulate magnetic properties of van der Waals materials.
\end{abstract}

\date{\today}

\maketitle

\section{Introduction}

Magnetic frustration in quantum systems represents one of the fundamental
ingredient to stabilize exotic magnetic order and, ultimately, 
quantum spin-liquid (QSL) states\cite{Balents2010,Lee2008,Broholm2020,RevModPhys.89.025003,Savary_2016}. 
Interest in quantum-spin-liquids has been fueled by their potential
emergent Majorana physics\cite{Kitaev2006}, 
high-temperature superconductivity\cite{ANDERSON1987,PhysRevX.6.041007},
and use for topological quantum computing\cite{PhysRevX.10.031014}. A variety of 
materials have been proposed as QSL candidates\cite{Han2012,Fu2015,Powell_2011, PhysRevX.9.031047,RevModPhys.88.041002, Takagi2019, PhysRevLett.91.107001, Yamashita2008,
PhysRevB.77.104413,PhysRevLett.112.177201, PhysRevLett.98.107204,PhysRevB.100.144432,Bordelon2019,PhysRevX.11.021044}, yet observing their QSL state is found to be highly sensitive to details of their Hamiltonian. While natural QSL materials are rare and require very precise fine-tuning, engineered materials allow to drastically overcome such challenges. Interestingly, van der Waals QSL candidates such as TaS$_2$\cite{Law2017,PhysRevResearch.2.013099,manasvalero2020multiple} and TaSe$_2$\cite{Chen2020} provide versatile platforms for a variety of quantum engineering methods such as straining\cite{Guinea2009,PhysRevLett.124.087205}, twisting\cite{PhysRevLett.99.256802,moire2021,PhysRevResearch.3.033276}, impurity engineering\cite{GonzalezHerrero2016,PhysRevLett.118.087203,PhysRevResearch.2.033466} as well as Coulomb engineering, that could potentially promote QSL behavior in these materials. 

\begin{figure}[t!]

\center
\includegraphics[width=\linewidth]{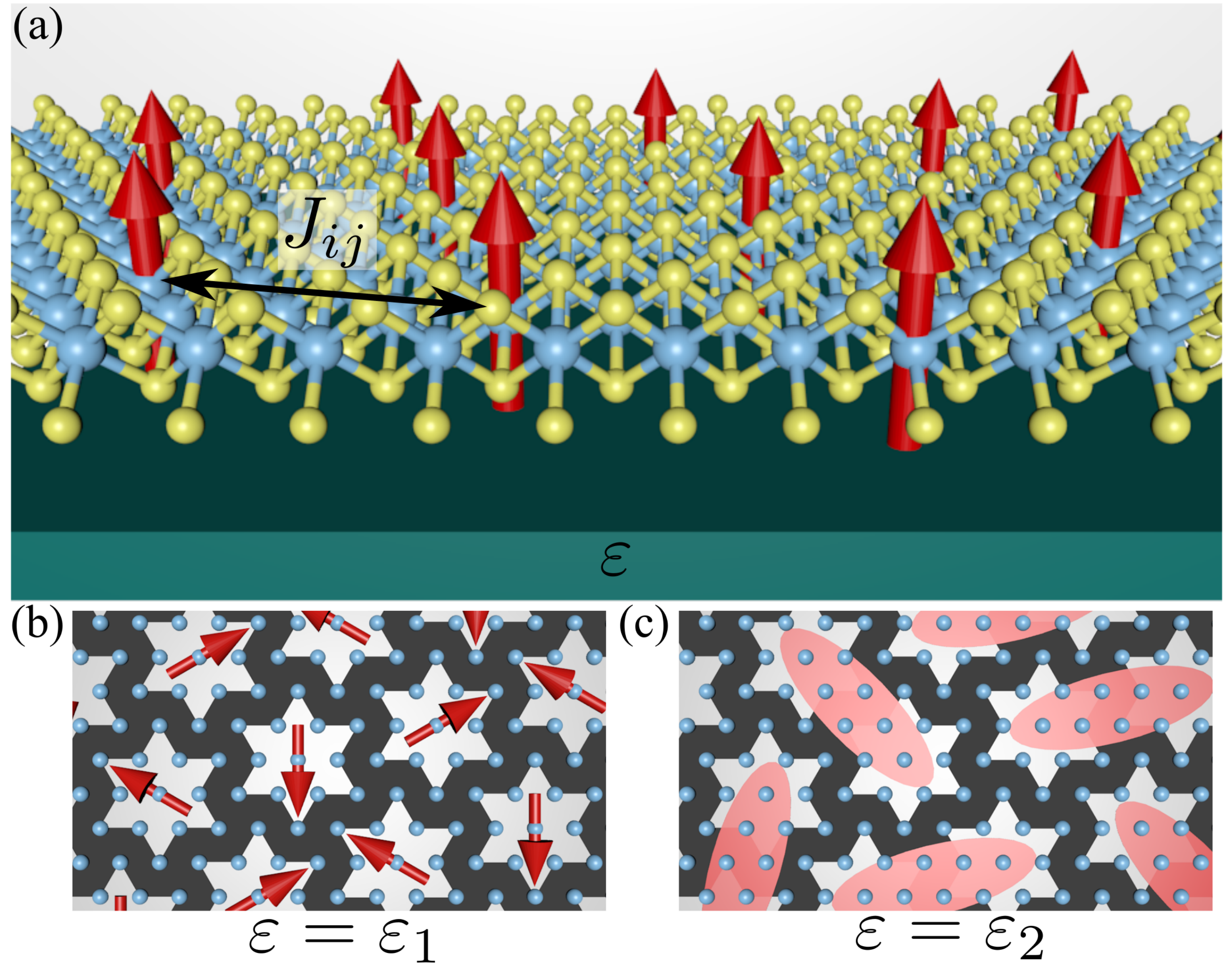}
\caption{ (a) Sketch of 1T-TaS$_2$ Coulomb engineered by a substrate with dielectric constant $\varepsilon$. The red arrows represent the local magnetic moments in the SOD unit cell. The superexchange couplings $J_{ij}$ between these magnetic moments can be tuned with Coulomb engineering. (b) Sketch of the helical spiral ground state in 1T-TaS$_2$. The red arrows represent the direction of the local magnetic moment. (c) Sketch of the QSL ground state in 1T-TaS$_2$.
}
\label{fig1}
\end{figure}

Coulomb engineering\cite{PhysRevLett.98.136805,PhysRevB.92.085102,Archana2017} refers to a strategy to tailor many-body interactions by means of dielectric environments, which is particularly efficient for low-dimensional materials. This is due to the pronounced role of non-local Coulomb interactions, which decisively define many-body properties in low-dimensional systems and which can be simultaneously efficiently externally modified. In this way band gaps\cite{Malte2016,Archana2017,steinhoff_exciton_2017,Iqbal2019,PhysRevB.102.115111,PhysRevLett.123.206403,vanloon2020coulomb} as well as excitonic\cite{Archana2017,PhysRevB.96.045431,PhysRevLett.123.206403} or plasmonic\cite{da_jornada_universal_2020,Jiang_2021} excitations and even topological properties\cite{PhysRevResearch.3.013265} can be precisely tailored in 2D and 1D systems with the help of (structured) dielectric substrates.
Furthermore, Coulomb interactions also play a crucial role in magnetic van der Waals materials, affecting the magnetic exchange between localized magnetic moments. However, up to date, controlling magnetic properties of van der Waals materials via Coulomb engineering has been limited to the example of CrI$_3$\cite{Soriano2021}.

Here, we put forward Coulomb engineering as a powerful tool to manipulate the ground states of magnetic van der Waals materials. 
Using the frustrated van der Waals magnet 1T-TaS$_2$ as a prototypical example, we show how substrate screening renormalizes its low-energy electronic dispersion and how this controls the internal magnetic superexchange interactions.
Together with direct exchange interactions, this allows to externally and non-invasively tailor the magnetic ground state in the material via changes to its dielectric environment.
Importantly, beyond the case of 1T-TaS$_2$ analyzed here, our proposal provides a starting point
towards the engineering of artificial magnets via tailored electronic interactions.
Ultimately, these results put forward the control of long-range Coulomb interactions as a versatile strategy for quantum matter design.

\section{Effective model for monolayer 1T-TaS$_2$}

We now elaborate on the material we will focus on in our discussion, 1T-TaS$_2$.
1T-TaS$_2$ hosts a charge-density-wave (CDW) instability leading to the formation of the Star-of-David (SOD) unit cell with 13 Ta atoms at low temperature\cite{Law2017,Cho2016,PhysRevX.7.041054,Kratochvilova2017,HF2021,Wang2020}. The CDW as well as spin-orbit coupling (SOC) result in a half-filled narrow band at the Fermi energy, with a bandwidth of a few 10meV\cite{PhysRevB.73.073106}. Together with sizable Coulomb interactions\cite{PhysRevB.105.L081106} this renders 1T-TaS$_2$ a correlated insulator rather than a simple metal. The electrons form local magnetic moments with $S=1/2$ at each SOD, and interact via exchange and superexchange coupling (Fig.~\ref{fig1}(a)), resulting in potential helical spiral and QSL ground states, as illustrated in Fig.~\ref{fig1}(b) and (c), respectively.

We capture the low energy physics of 1T-TaS$_2$ with a single Wannier orbital
model including long-range electronic interactions:
\beq \label{eq1}
\begin{aligned}
H=&\sum_{i,j,\sigma}t_{ij}c^\dag_{i,\sigma}c_{j,\sigma}+U\sum_in_{i,\uparrow}n_{i,\downarrow}\\&+\sum_{i,j,\sigma,\sigma'}\frac{V_{ij}}{2}n_{i,\sigma}n_{j,\sigma'},
\end{aligned}
\eeq
where $\sigma$ and $\sigma'$ are spin indices, and $n_{i,\sigma}=c^\dag_{i,\sigma}c_{i,\sigma}$. The hoppings $t_{ij}$ are fitted to DFT data\cite{PhysRevB.105.L081106} where we find hoppings up to the 14$^\text{th}$ neighbor. The hoppings exhibit an oscillating behavior in addition to a decay with distance. This oscillating behavior stems from the nature of the Wannier wavefunctions of this material, and is inherited from the electronic structure from first principles calculations.
We parameterize the long-range Coulomb interactions $V_{ij}$ via a modified Yukawa potential of the form
\beq \label{eq2}
V_{ij}=\frac{U}{\sqrt{1+\left(\frac{4\pi\varepsilon_0Ur_{ij}}{e^2}\right)^2}}e^{-r_{ij}/r_\text{TF}},
\eeq
where $U$ is the on-site Coulomb interaction and $r_{ij}$ the distance between sites $i$ and $j$. The included Ohno potential\cite{Ohno1964,vanloon2020coulomb} results in a $r^{-1}$ long-wavelength behavior, which is further suppressed by the exponential term controlled by an effective screening length $r_\text{TF}$. This way, the non-local Coulomb interaction is fully parameterized by the local interaction $U$ and the screening length $r_\text{TF}$. It is worth noting that environmental screening to layered materials, such as resulting from dielectric substrates, is strongly non-local\cite{Keldysh,PhysRevLett.98.136805,PhysRevB.92.085102} such that long-ranged interactions $V_{ij}$ are stronger reduced than the local one $U$. To fully characterize this model we, however, treat $U$ and $r_\text{TF}$ as independent parameters in the following, understanding that any environmental screening will reduce both simultaneously.

The interacting model of Eq.~\eqref{eq1} is analyzed in two steps.
In the limit $U\gg V_{ij}$, we can first integrate out the
long-range interactions $V_{ij}$, leading to a renormalized
dispersion for the low-energy band\cite{intVeld2019,PhysRevB.95.245130}.
The resulting Hamiltonian $\tilde{H}$ takes the form 
\beq \label{eq3}
\tilde{H}=\sum_{i,j,\sigma}\tilde{t}_{ij}c^\dag_{i,\sigma}c_{j,\sigma}+U\sum_in_{i,\uparrow}n_{i,\downarrow},
\eeq
where $\tilde{t}_{ij}$ are the renormalized hoppings derived from Eq.~\eqref{eq1} using a Hatree-Fock variational wavefunction enforcing time-reversal symmetry.

\begin{figure}[t!]
\center
\includegraphics[width=\linewidth]{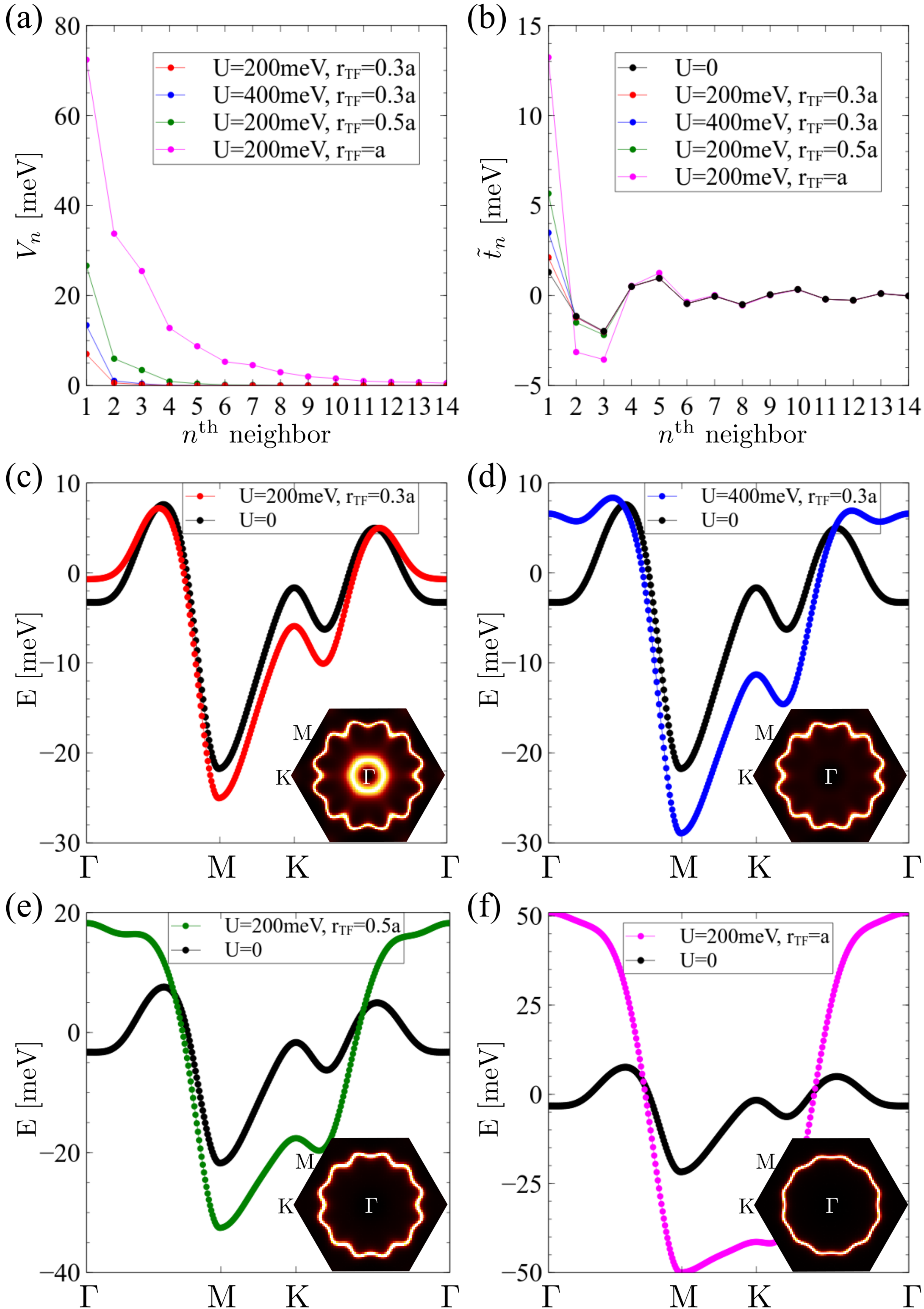}
\caption{(a) Long-range Coulomb interactions $V_{n}$ and (b) the corresponding renormalized hoppings $\tilde{t}_{n}$ for different values of parameters $U$ and $r_{\text{TF}}$. (c-f) band structure renormalization for different values of $U$ and $r_{\text{TF}}$ in (a). The color matches with the color used in (a). The insets show the corresponding Fermi surfaces for the renormalized bands.}
\label{fig2}
\end{figure}

For $U$ between $100$ and $500\,$meV as estimated for 1T metallic TMDCs \cite{PhysRevB.105.L081106,Kamil2018,Pizarro2020} and several choices of $r_\text{TF}$, the corresponding long-range Coulomb interactions $V_n$ and renormalized hoppings $\tilde{t}_n$ are shown in Figs.~\ref{fig2}(a,b), where $V_n$ denotes $n_\text{th}$ neighbor Coulomb interaction and similarly for $\tilde{t}_n$. We see that $V_n$ is increased by increasing $U$ and $r_\text{TF}$, which mostly affects $\tilde{t}_1$ for $r_\text{TF} < a$ ($a$ is the CDW lattice constant), while the long-range interaction with $r_\text{TF} = a$ also modulates hoppings up to $\tilde{t}_5$. In Figs.~\ref{fig2}(c-f) we show the bare electronic dispersion after integrating out long-range Coulomb interactions $V_n$, corresponding to the band structure of $\tilde{H}_\text{bare}=\sum_{i,j,\sigma}\tilde{t}_{ij}c^\dag_{i,\sigma}c_{j,\sigma}$. 
For $r_\text{TF} = 0.3a$ (with $V_{n>1} \approx 0$) the renormalized bare bandwidth is not significantly affected and we find only modifications to the dispersion around $\Gamma$ upon changing $U$, c.f. Figs.~\ref{fig2}(c,d). Increasing $r_\text{TF} > 0.3a$ yields $V_{n>1} > 0$, which decisively affects both, the renormalized bare bandwidth and the overall dispersion, c.f. Figs.~\ref{fig2}(e,f). As maintaining the Mott regime requires that the bandwidth renormalization should not be too large, we focus on $r_\text{TF}\leq 0.4a$ in the following. In addition to the bandwidth renormalization, we observe a Coulomb controlled Lifshitz transition: as $U$ and $r_\text{TF}$ increase, the electron pocket at $\Gamma$ vanishes, as depicted in the insets of Figs.~\ref{fig2}(c-f). This transition results in different dependencies of superexchange couplings on the Coulomb interactions as we discuss in the following. We note that the renormalized bare dispersions shown in Fig.~\ref{fig2}(c-f) do not explicitly take the local Coulomb interaction $U$ into account as in the Hamiltonian $\tilde{H}$ from Eq.~\eqref{eq3}. To obtain the electronic dispersion of $\tilde{H}$ including the effects of $U$, dynamical mean field theory (DMFT)\cite{Kamil2018,vanloon2020coulomb,PhysRevB.105.L081106,Pizarro2020} calculations are required. We show that this is not needed to analyze the magnetic properties of $\tilde{H}$ in the following.

\section{Coulomb engineering of the spin Hamiltonian of 1T-TaS$_2$}

\begin{figure}[t!]
\center
\includegraphics[width=\linewidth]{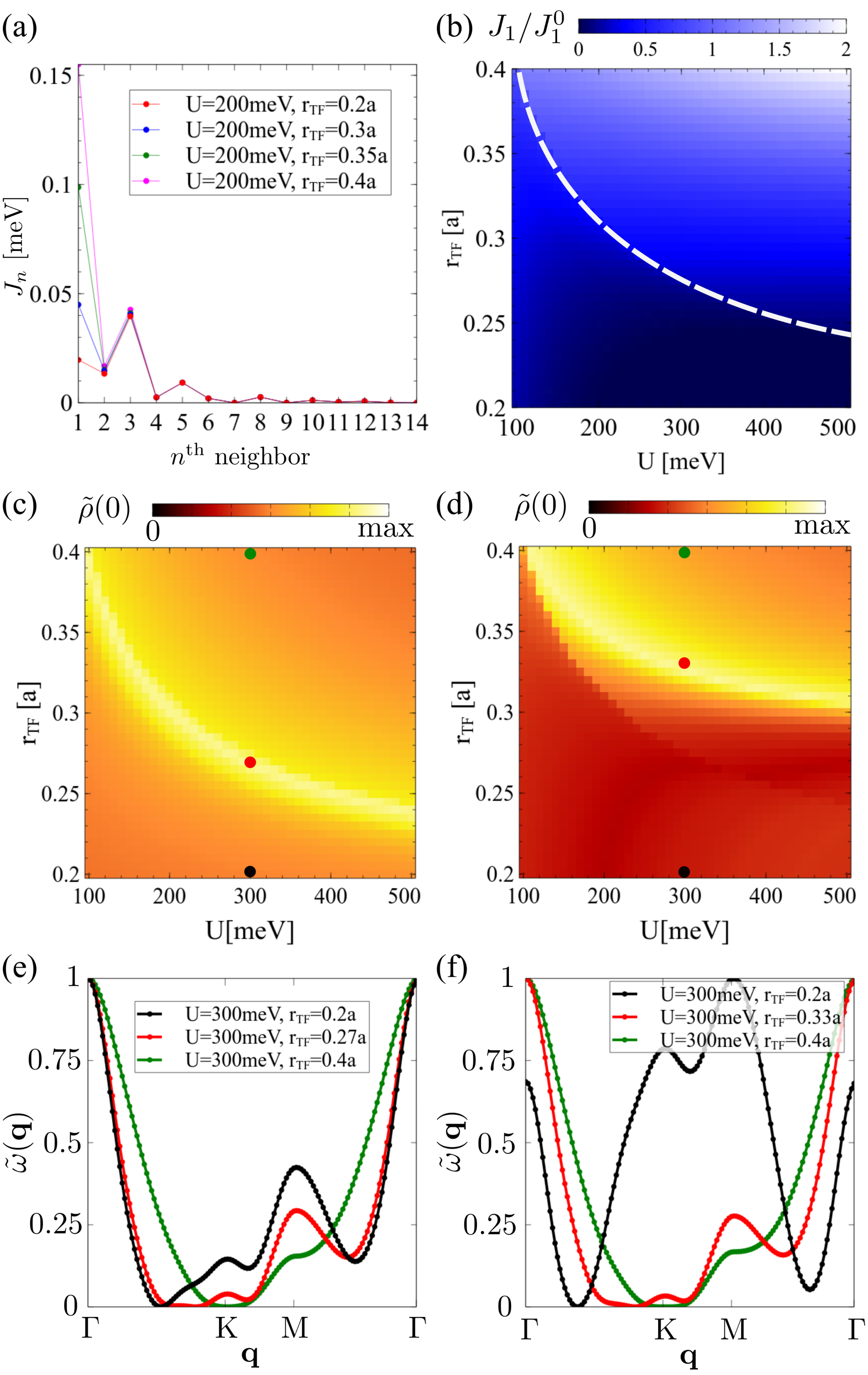}
\caption{(a) Long-range spin superexchange couplings $J_n$ for different values of $U$ and $r_{\text{TF}}$. Only $J_1$ is significantly influenced by $U$ and $r_{\text{TF}}$. (b) $J_1$ for different values of $U$ and $r_{\text{TF}}$, in units of $J_1^0=0.1$meV. The dashed line indicates a discontinuity stemming from the Lifshitz transition. (c,d) Ground state degeneracy $\tilde{\rho}(0)$ for our model Eq.\eqref{eq1}, as a function of $U$ and $r_\text{TF}$. In (d) we include a direct exchange of $J_\text{DX}=0.5J_1^0$. (e,f) The corresponding $\tilde{\omega}(\mathbf{q})$ for the points indicated in (c,d), respectively.}
\label{fig3}
\end{figure}

We now analyze the Hamiltonian from Eq.~\eqref{eq3} in the strong coupling limit, i.e. $U\gg \tilde t_{ij}$, using the Schrieffer-Wolf transformation, leading to an effective model for spin-degrees of freedom:
\beq \label{eq4}
\mathcal{H}=\sum_{i,j}J_{ij}\mathbf{S}_i\cdot\mathbf{S}_j,
\eeq
with $J_{ij}=2\frac{\tilde{t}_{ij}^2}{U}$. Due to the renormalization of $t_{ij}$ by $V_{ij}$, the magnetic superexchange interactions $J_{ij}$ are controlled by changes to $U$ and $r_\text{TF}$ as well.
In particular, we show in Fig.~\ref{fig3}(a) the renormalized $n^\text{th}$ neighbor exchange $J_n$. We see that $J_1$ exhibits a strong dependence on $U$ and $r_\text{TF}$, stemming from the significant renormalization of $\tilde{t}_1$. In Fig.~\ref{fig3}(b) we depict the full $U$ and $r_\text{TF}$ dependencies of $J_1$ in units of $J_1^0=0.1$meV, where $J_1^0$ is the typical magnitude of $J_1$ in our regime. We find two regimes [separated by the white dashed line in Fig.~\ref{fig3}(b)] stemming from the Coulomb driven
Lifshitz transition of the Hamiltonian of Eq.~\eqref{eq1}. In the upper regime $V_n$ has a long-range character yielding renormalized band structures with unoccupied and rather flat dispersions around $\Gamma$, while in the lower regime the Coulomb interaction is short-ranged ($U$ and $V_1$) and we find occupied electron pockets (with positive effective mass) around $\Gamma$. $J_1$ is correspondingly larger in the upper regime and strongly dependent on $r_\text{TF}$. In the lower regime $J_1$ is reduced and mostly dependent on $U$.

The model of Eq.~\eqref{eq4} thus realizes a long-range Heisenberg model with tunable frustration controlled by the local and non-local Coulomb interactions. While a full calculation of the phase diagram would require exactly solving the two-dimensional quantum many-body model, e.g. with tensor-network\cite{Yan2011,PhysRevLett.123.207203} or neural network quantum states\cite{Carleo2017,PhysRevB.100.125124}, here we will focus on highlighting the physical regime where the frustration of the system is maximal. A frustrated magnetic system has competing magnetic exchange interactions whose energies cannot be simultaneously minimized by any magnetic configuration\cite{PhysRevLett.107.260602}. As a consequence, classical magnetic ground states of a frustrated magnetic system exhibit a high degree of degeneracy, preventing the system from magnetic ordering and promoting a QSL state. We may thus characterize the frustration of a magnetic system by the degeneracy of classical magnetic ground states. We consider different non-collinear magnetic configurations characterized by a vector $\mathbf{q}$\cite{Egger2012,Chen_2022} and compute their energy $\omega(\mathbf{q})$ from the extended Hubbard model Eq. \eqref{eq1}. The resulting $\omega(\mathbf{q})$ is then shifted and scaled: $\tilde{\omega}(\mathbf{q})=(\omega(\mathbf{q})-\omega_0)/(\omega_1-\omega_0)$ with $\omega_0=\min(\omega(\mathbf{q}))$ and $\omega_1=\max(\omega(\mathbf{q}))$. The shifted ground state energy is 0, and the scaling allows comparison between systems with different magnitudes of magnetic exchange interactions. Finally, the degeneracy of the states at a given shifted and scaled energy $\omega$ can be given by
\beq \label{eq5}
\tilde{\rho}(\omega)=\int_{\mathbf{q}\in\text{BZ}}\frac{\text{d}^2\mathbf{q}}{(2\pi)^2}\delta(\tilde{\omega}(\mathbf{q})-\omega).
\eeq
In particular, $\tilde{\rho}(0)$ characterizes the ground state degeneracy of a magnetic system, and a more frustrated system is characterized by larger $\tilde{\rho}(0)$.

The computation of $\omega(\mathbf{q})$ can be done by solving Eq.~\eqref{eq1} self-consistently assuming the solution of a non-collinear magnetic state characterized by the vector $\mathbf{q}$, where without loss of generality we take the magnetization to be in $y$-$z$ plane. Alternatively, we can perform a unitary transformation that aligns the magnetization on every site to $z$ axis\cite{Egger2012,Braunecker2010,Chen_2022}: $H_\mathbf{q}=U_\mathbf{q}^\dag H U_\mathbf{q}$ with $U_\mathbf{q}= \prod_i e^{-\frac{i}{2}\mathbf{q}\cdot\mathbf{r}_i\sigma_{x,i}}$, and solve it self-consistently assuming ferromagnetism in $z$ direction. The transformed $H_\mathbf{q}$ is:
\beq \label{eq6}
\begin{aligned}
H_\mathbf{q}=&\sum_{i,j,\sigma,\sigma'}t_{ij,\sigma\sigma'}(\mathbf{q})c^\dag_{i,\sigma}c_{j,\sigma'}+U\sum_in_{i,\uparrow}n_{i,\downarrow}\\&+\sum_{i,j,\sigma,\sigma'}\frac{V_{ij}}{2}n_{i,\sigma}n_{j,\sigma'},
\end{aligned}
\eeq
where $t_{ij,\sigma\sigma'}(\mathbf{q})=t_{ij}e^{\frac{i}{2}\mathbf{q}\cdot(\mathbf{r}_i-\mathbf{r}_j)(\sigma_x)_{\sigma\sigma'}}$. In the limit $U\gg V_{ij}$, we can first integrate out $V_{ij}$ as we did for Eq.~\eqref{eq1}, leading to
\beq \label{eq7}
\tilde{H}_\mathbf{q}=\sum_{i,j,\sigma,\sigma'}\tilde{t}_{ij,\sigma\sigma'}(\mathbf{q})c^\dag_{i,\sigma}c_{j,\sigma'}+U\sum_in_{i,\uparrow}n_{i,\downarrow},
\eeq
where $\tilde{t}_{ij,\sigma\sigma'}(\mathbf{q})=\tilde{t}_{ij}e^{\frac{i}{2}\mathbf{q}\cdot(\mathbf{r}_i-\mathbf{r}_j)(\sigma_x)_{\sigma\sigma'}}$
\footnote{We can perform a change of basis to Eq.~\eqref{eq6}: $d^\dag_{i,\uparrow}=\cos\theta c^\dag_{i,\uparrow}+i\sin\theta c^\dag_{i,\downarrow}$ and $d^\dag_{i,\downarrow}=i\sin\theta c^\dag_{i,\uparrow}+\cos\theta c^\dag_{i,\downarrow}$. This does not change the $V_{ij}$ term, and the Hamiltonian becomes Eq.~\ref{eq1} with only changes in the $U$ term. The resulting $\tilde{t}_{ij}$ after integrating out $V_{ij}$ is thus given by Eq.~\ref{eq3}. Transforming back to the original basis results in Eq.~\eqref{eq7}.}. With the assumption of ferromagnetism along $z$ direction: $\langle n_{i,\uparrow}\rangle=1, \langle n_{i,\downarrow}\rangle=0$, Eq.~\eqref{eq7} reduces to
\beq \label{eq8}
H^\text{MF}_\mathbf{q}=\sum_{i,j,\sigma,\sigma'}\tilde{t}_{ij,\sigma\sigma'}(\mathbf{q})c^\dag_{i,\sigma}c_{j,\sigma'}+\frac{U}{2}\sum_i(n_{i,\downarrow}-n_{i,\uparrow})
\eeq
up to a constant, and $\omega(\mathbf{q})$ can be computed by summing over occupied states in Eq.~\eqref{eq8}.

Now we move on to compute $\tilde{\rho}(0)$ for our model with different $U$ and $r_{\text{TF}}$. We find that, $\tilde{\rho}(0)$ takes large values close to the Lifshitz transition and smaller values at other places (Fig.\ref{fig3}(c)). In particular, below the Lifshitz transition, $\tilde{\omega}(\mathbf{q})$ exhibits a minima between $\Gamma$ and K (Fig.\ref{fig3}(e)). As we approach the Lifshitz transition, $J_1$ increases, lowering $\tilde{\omega}(\mathbf{q})$ around $\mathbf{q}=\text{K}$, eventually this results in an almost flat dispersion of $\tilde{\omega}(\mathbf{q})$ in the region between $\Gamma$ and K. Above the Lifshitz transition, $J_1$ further increases and starts to dominate other exchanges. This results in the stabilization of the helical spiral state with $\mathbf{q}=\text{K}$, where $\tilde{\omega}(\mathbf{q})$ takes its minimum. We have thus found that by tuning the substrate-screening of 1T-TaS$_2$, we can enhance the magnetic frustration and make the QSL ground state more favorable.

We note that the extended Hubbard model Eq.~\eqref{eq1} and the Heisenberg model Eq.~\eqref{eq4} consider only superexchange couplings stemming from the local repulsion. Apart from that, it is important to note that direct exchange stemming from the overlap of Wannier centers would also appear in the effective Heisenberg model\cite{PhysRevB.105.L081106,PhysRevB.94.214411}. In particular, strong direct exchange interactions promote a ferromagnetic ground state in 1T-TaS$_2$\cite{PhysRevB.105.L081106}. To account for this, we now include the direct exchange term $J_\text{DX}\sum_{\langle i,j\rangle}\mathbf{S}_i\cdot\mathbf{S}_j$, which results from Hund's exchange interaction and can thus be assumed to be independent of the environmental screening~\cite{Soriano2021}. The existence of a finite direct exchange merely cancels with the superexchange $J_1$. Thus the most frustrated regime appears at larger $J_1$, which for fixed $U$ appears at larger $r_\text{TF}$ (Fig.\ref{fig3}(d)). For smaller $r_\text{TF}$, $\tilde{\omega}(\mathbf{q})$ has a minimum along $\Gamma$-K (Fig.\ref{fig3}(f)), which would approach $\Gamma$ for sufficiently small $r_\text{TF}$ or sufficiently large $J_\text{DX}$. Thus we find that the existence of finite direct exchange does not influence our main result.
We have thus demonstrated the potential of Coulomb engineering to tailor the magnetic ground state and the magnetic frustration. In practice this can be achieved by exposing the monolayer to different substrates, such as SiO$_2$, SrTiO$_3$, or hBN, or by embedding it in a tunable dielectric environment\cite{PhysRevB.101.121110,Saito2020}, where the effective dielectric constant can be controlled by the thickness of the substrate and by gating the substrate. 
 
\section{Discussion}

Finally, we comment on several aspects that should be addressed in future work. First, the exact dependence of the parameters $U$ and $r_\text{TF}$ on the substrate dielectric constant can be solved with ab initio calculations in specific experimental setups. Second, our analysis focuses on a spin-isotropic model, where potentially anisotropic terms stemming from spin-orbit coupling are not included, which is motivated by recent DFT study\cite{PhysRevB.105.L081106}. The inclusion of spin-orbit coupling in the low-energy model would give rise to spin-dependent hoppings, which in turn could induce anisotropic exchange in the spin model generating an even richer phase diagram. Third, we used a Hatree-Fock variational wavefunction to capture the bandwidth renormalization induced by $V_{ij}$, while a more accurate approach would be extensions of the Peierls-Feynman-Bogoliubov variational principle \cite{intVeld2019,PhysRevLett.111.036601,PhysRevB.94.165141} taking the full non-local Coulomb and nearest-neighbor exchange interactions into account. Yet for long-range $V_{ij}$ up to 3$^\text{rd}$ neighbor, such an approach involves heavy computation that would go beyond the scope of this work. Fourth, we note that a more sophisticated description of the frustration can be performed by exactly solving the many-body system and analyzing the mathematical structure of the associated reduced density matrices\cite{PhysRevLett.107.260602}. 
Fifth, our analysis focused on the bilinear spin interactions of the spin-Hamiltonian, yet at intermediate interactions the  strength biquadratic and four-spin ring-exchange couplings can have sizable contributions\cite{pasquier2021textitab}.
Finally, although we used the 1T-TaS$_2$ band structure as an example here, the same methodology can be applied to other van der Waals materials such as 1T-NbSe$_2$\cite{PhysRevB.98.045114,Kamil2018,Yuki2016,Liu2021}, 1T-NbS$_2$\cite{PhysRevB.102.155115,PhysRevB.105.035119}, 1T-TaSe$_2$\cite{Ruan2021}, as well as other 1T-dichalcogenide alloys, and potentially twisted graphene multilayers\cite{PhysRevB.100.161102,PhysRevLett.123.096802}, where non-local Coulomb interactions are not negligible. As an outlook, Coulomb engineering can also be performed with spatially structured\cite{Malte2016,Archana2017,PhysRevB.102.115111,PhysRevResearch.3.013265,Jiang_2021} or anisotropic\cite{Xia2014,PhysRevLett.123.216403} screening environments giving rise to spatially dependent exchange interactions, that potentially leads to coexisting ground states of different character within the same, homogeneous layered material. 

\section{Conclusions}

To summarize, we have put forward a strategy to control the magnetic ground state of strongly correlated layered materials, which shows the potential to drive a two-dimensional magnetic material to a tunable frustrated regime by means of Coulomb engineering. Taking as a starting point the Wannier model for the nearly flat band structure in 1T-TaS$_2$, we demonstrated that tunable screening drastically impacts the low-energy spin model of the system. In particular, we showed that the long-range Coulomb interactions result in bandwidth renormalization at the Hatree-Fock level. The renormalized bandwidth, together with the long-range nature of the Wannier model gives rise in the strongly interacting limit to a screening-dependent frustrated Heisenberg model. Finally, we showed how tuning the long-range Coulomb interaction via screening can bring the system to a highly degenerate regime by analyzing the frustration of the spin system. Our proposal demonstrates how substrate-dependent screening, widely present in studies of van der Waals heterostructures, provides a powerful strategy to stabilize unconventional correlated states of matter. Ultimately, our results provide a starting point towards tailoring frustrated quantum magnetism via Coulomb engineering, potentially allowing to stabilize tunable frustrated states in a variety of van der Waals magnets.\\

\textbf{Acknowledgements}
We acknowledge
the computational resources provided by
the Aalto Science-IT project,
and the
financial support from the
Academy of Finland Projects No.
331342 and No. 336243.
We thank M. I. Katsnelson, A. N. Rudenko, A. A. Bagrov, T. Westerhout,
P. Liljeroth and V. Va\v{n}o for fruitful discussions.

\begin{appendix}

\end{appendix}

\bibliography{biblio}{}

\end{document}